\pdfoutput=1

\documentclass[10pt,aps,prl,twocolumn,superscriptaddress]{revtex4-1}
\pdfoutput=1
\usepackage{graphicx}
\usepackage{amssymb}
\usepackage{moresize}
\usepackage{amsmath}
\usepackage{bm}
\usepackage{txfonts}
\usepackage{letltxmacro}
\LetLtxMacro{\oldsqrt}{\sqrt}
\renewcommand{\sqrt}[2][\mkern8mu]{\mkern-4mu\mathop{\oldsqrt[#1]{#2}}}

\usepackage[usenames]{color}

\def\be{\begin{equation}}
\def\ee{\end{equation}}
\def\bea{\begin{eqnarray}}
\def\eea{\end{eqnarray}}
\def\ba{\begin{array}}
\def\ea{\end{array}}

\begin{document}

\title{Scaling theory for mechanical critical behavior in fiber networks}


\author{Jordan Shivers}
\affiliation{Department of Chemical and Biomolecular Engineering, Rice University, Houston, TX 77005, USA}
\affiliation{Center for Theoretical Biological Physics, Rice University, Houston, TX 77030, USA}
\author{Sadjad Arzash}
\affiliation{Department of Chemical and Biomolecular Engineering, Rice University, Houston, TX 77005, USA}
\affiliation{Center for Theoretical Biological Physics, Rice University, Houston, TX 77030, USA}
\author{Abhinav Sharma}
\affiliation{Leibniz Institute of Polymer Research Dresden, Dresden, Germany}
\author{Fred C.\ MacKintosh}
\affiliation{Department of Chemical and Biomolecular Engineering, Rice University, Houston, TX 77005, USA}
\affiliation{Center for Theoretical Biological Physics, Rice University, Houston, TX 77030, USA}
\affiliation{Departments of Chemistry and Physics \& Astronomy, Rice University, Houston, TX 77005, USA}



\begin{abstract}

As a function of connectivity, spring networks exhibit a critical transition between floppy and rigid phases at an isostatic threshold.
For connectivity below this threshold, fiber networks were recently shown theoretically to exhibit a rigidity transition with corresponding critical signatures as a function of strain. Experimental collagen networks were also shown to be consistent with these predictions. We develop a scaling theory for this strain-controlled transition. Using a real-space renormalization approach, we determine relations between the critical exponents governing the transition, which we verify for the strain-controlled transition using numerical simulations of both triangular lattice-based and packing-derived fiber networks.

\end{abstract}
 
\maketitle

It has long been recognized that varying connectivity can lead to a rigidity transition in networks such as those formed by springlike, central force (CF) connections between nodes. 
Maxwell introduced a counting argument for the onset of rigidity for such systems in $d$ dimensions with $N$ nodes, in which the number of degrees of freedom $dN$ is balanced by the number of constraints $Nz/2$, where $z$ is the average coordination number of the network \cite{Maxwell1864}. 
The transition at this \emph{isostatic} point of marginal stability has been shown to exhibit signatures of criticality. 
Such a balance of constraints and degrees of freedom is important for understanding rigidity percolation and jamming \cite{thorpe1983continuous,feng1984percolation,cates1998jamming,Liu1998,VanHecke2010}.
Even in networks with additional interactions that lead to stability below the CF isostatic point, 
the mechanical response can still exhibit strong signatures of criticality in the vicinity of the CF isostatic point \cite{Wyart2008,Broedersz2011NatPhys,Das2012,Feng2016}. 
More recently, criticality has been shown in fiber networks as a function of strain for systems well below the isostatic point \cite{sharma2016strain}. 

While both jammed particle packings and fiber networks exhibit athermal ($T=0$) mechanical phase transitions and superficially similar critical behavior, these systems have strong qualitative differences. There is growing evidence that the jamming transition  is mean-field \cite{VanHecke2010,Goodrich30082016}, while fiber networks to date have shown non-mean-field behavior \cite{Broedersz2011NatPhys,Das2012,Feng2016,sharma2016strain}. 
Goodrich et al. recently proposed a scaling theory and performed numerical simulations of jamming to both demonstrate mean-field exponents and support the conclusion that the upper critical dimension $d_u=2$ for the jamming transition \cite{Goodrich30082016}. 
Although many aspects of the critical behavior of fiber networks, including various critical exponents, have been quantified, a theory has been lacking to identify critical exponents or even scaling relations among exponents. 
Here, we develop a scaling theory for both the sub-isostatic, strain controlled transition, as well as the transition in $z$ near the isostatic point for athermal fiber networks. We derive scaling relations among the various exponents and demonstrate good agreement with numerical simulations. 
Interestingly, our results imply that the upper critical dimension for fiber networks is $d_u>2$, in contrast with jamming packings. 

\begin{figure}[b]
\includegraphics[width=0.8\columnwidth]{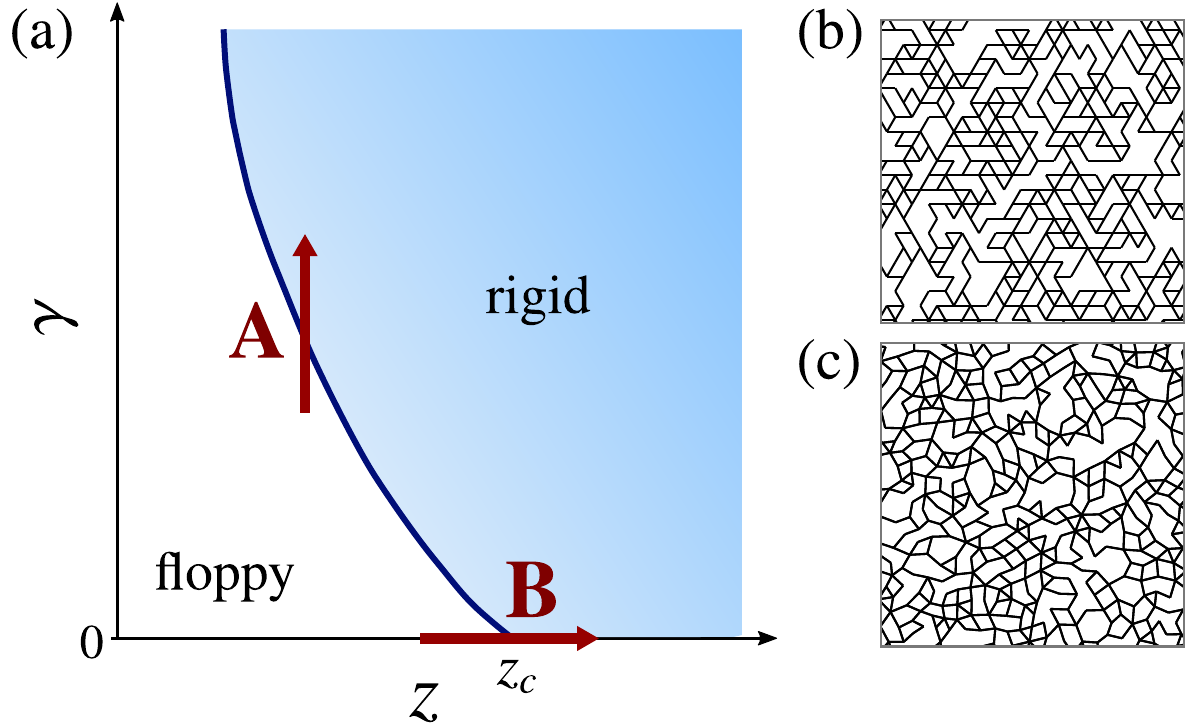}
\caption{\label{fig1}  (a)  Schematic phase diagram depicting the state of mechanical rigidity of a central force network as a function of coordination number $z$ and applied shear strain $\gamma$. The arrow $A$ depicts the strain-controlled transition and $B$ depicts the transition at the isostatic point. With the addition of bending interactions, the floppy region becomes instead bending-dominated, but the critical curve $\gamma_c(z)$ vs. $z$ remains the same. (b) Portion of a a triangular network and (c) a 2D packing-derived network, both diluted to $z = 3.3<z_c$.
}
\end{figure}

Near the isostatic point with average coordination number $z=z_c=2d$, spring networks exhibit shear linear moduli $G$ that vary as a power of $|z-z_c|$ for $z>z_c$ \cite{Ellenbroek2006,VanHecke2010,Wyart2008}.
In the floppy or sub-isostatic regime with $z<z_c$, such systems can be stabilized by introducing additional interactions \cite{Wyart2008,Broedersz2011NatPhys} or by imposing stress or strain \cite{Alexander1998,Sheinman2012}.
It was recently shown that sub-isostatic networks undergo a transition from floppy to rigid as a function of shear strain $\gamma$ \cite{sharma2016strain,Sharma2016}. Moreover, this fundamentally nonlinear transition was identified as a line of critical points characterized by a $z-$dependent threshold $\gamma_c(z)$, as sketched in Fig.\ \ref{fig1}a.
Above this strain threshold, the differential or tangent shear modulus $K=d\sigma/d\gamma$ scales as a power law in strain, with $K\sim\left|\gamma-\gamma_c\right|^f$.  Introducing bending interactions with rigidity $\kappa$ between nearest neighbor bonds stabilizes sub-isostatic networks below the critical strain, leading to $K\sim\kappa$ for $\gamma < \gamma_c$.  Both of these regimes are captured by the scaling form \cite{Widom1965}
\be
K\approx\left|\gamma-\gamma_c\right|^f\mathcal{G}_{\pm}\left({\kappa}/{\left|\gamma-\gamma_c\right|^\phi}\right)\label{Widom}
\ee
in which the branches of the scaling function $\mathcal{G}_{\pm}$ account for the strain regimes above and below $\gamma_c$.
This scaling form was also shown to successfully capture the nonlinear strain stiffening behavior observed in experiments on reconstituted networks of collagen, a filamentous protein that provides mechanical rigidity to tissues as the primary structural component of the extracellular matrix \cite{Sharma2016}. Collagen constitutes an excellent experimental model system on which to study this transition, as it forms elastic networks that are deeply sub-isostatic ($z\approx3.4$ \cite{Lindstrom2013}, whereas $z_c=6$ in 3D) in which individual fibrils have sufficiently high bending moduli to be treated as athermal elastic rods.

\textit{Scaling theory} \textemdash\, For the strain-controlled transition at a fixed $z<z_c$ (arrow $A$ in Fig.\ \ref{fig1}a), we define a reduced variable $t=\gamma-\gamma_c$ that vanishes at the transition and  
let $h(t,\kappa)$ denote the Hamiltonian or elastic energy per unit cell. This energy depends on the bending stiffness $\kappa$ that also vanishes at the transition. 
Assuming the system becomes critical as $t, \kappa\rightarrow0$, we consider the real-space renormalization of the system when scaled by a factor $L$ to form a block or effective unit cell composed of $L^d$ original cells, where $d$ is the dimensionality of the system \cite{Kadanoff}. 
Under this transformation, the energy per block becomes 
$h(t',\kappa')=L^d h(t,\kappa)$,
where $t'$ and $\kappa'$ are renormalized values of the respective parameters. 
We assume that the parameters evolve according to
$t\rightarrow t'=tL^x$ and $\kappa\rightarrow\kappa'=\kappa L^y$,
where the exponents $x,y$ can be assumed to be positive, since the system must evolve away from criticality. 
Combining these, we find the elastic energy 
\be
h(t,\kappa)=L^{-d}h(tL^x,\kappa L^y).\label{ScalingGamma}
\ee 

The stress is simply given by the derivative with respect to strain of the elastic energy per volume, which is proportional to $h(t,\kappa)$. Thus,
$
\sigma\sim\frac{\partial}{\partial\gamma}h\sim\frac{\partial}{\partial t}h(t,\kappa)\sim L^{-d+x}h_{1,0}(tL^x,\kappa L^y)
$
and the stiffness 
\be
K=\frac{\partial}{\partial\gamma}\sigma\sim\frac{\partial^2}{\partial t^2}h(t,\kappa)\sim L^{-d+2x}h_{2,0}(tL^x,\kappa L^y),\label{Kdef}
\ee
where $h_{n,m}$ refers to the combined $n$-th partial with respect to $t$ and $m$-th partial with respect to $\kappa$ of $h$. Being derivatives of the energy with respect to the control variable $\gamma$, the stress and stiffness are analogous to the entropy and heat capacity for a thermal system with phase transition controlled by temperature.
If we let $L=|t|^{-1/x}$, then the correlation length scales according to $\xi\sim L\sim|t|^{-\nu}$, from which we can identify the correlation length exponent $\nu=1/x$. Thus, the stiffness can be expressed as in Eq.\ \eqref{Widom}, where
$\mathcal{G}_{\pm}\left(s\right)
\sim h_{2,0}(\pm 1,s)$ and
\be
f=d\nu-2\quad\mbox{and}\quad\phi=y\nu. \label{strain_exponents}
\ee
The first of these is a hyperscaling relation analogous to that for the heat capacity exponent for thermal critical phenomena, but in which $f>0$ corresponds to nonlinear stiffness $K$ that is continuous.
For $\gamma>\gamma_c$, we expect that $h_{2,0}(1,s)$ is approximately constant for $s\ll1$, so that $K\sim|\gamma-\gamma_c|^f$, while
for $\gamma<\gamma_c$ we expect that 
$h_{2,0}(-1,s)\sim s$ for $s\ll1$, 
so that 
\be
K\sim\kappa|\gamma-\gamma_c|^{-\lambda},\label{susceptibility}
\ee 
consistent with a bend-dominated elastic regime. Moreover, the susceptibility-like exponent is expected to be $\lambda=\phi-f$. 

Near the critical strain, networks exhibit large, nonaffine internal rearrangements in response to small changes in applied strain \cite{Sharma2016, sharma2016strain}.  In the absence of thermal fluctuations in such athermal networks, these nonaffine strain fluctuations are analogous to divergent fluctuations in other critical systems. 
In response to an incremental strain strain step $\delta \gamma$, the nonaffine displacement of the nodes is expected to be proportional to $\delta \gamma$.
Thus, the nonaffine fluctuations can be captured by $\delta\Gamma\sim\langle\left|\delta \mathbf{u}-\delta \mathbf{u}^{A}\right|^2\rangle/\delta\gamma^2$, 
where $\delta \mathbf{u}-\delta \mathbf{u}^{A}$ represents the deviation relative to a purely affine displacement $\delta \mathbf{u}^{A}$. For large systems with small $\kappa$, $\delta\Gamma$ diverges as $t\rightarrow0$ \cite{Sharma2016}.  
This divergence can also be derived from the energy $h(t,\kappa)$ in the limit of small bending stiffness $\kappa$, as follows. For small bending energy, the nonaffine displacements $\delta u^2$ are determined by the minimization of $h$, which should then be due to purely bending: $h\sim\kappa\delta u^2\sim\kappa\delta\gamma^2\delta\Gamma$. Thus, the nonaffine fluctuations are predicted to diverge as
\be
\delta\Gamma\sim\frac{\partial}{\partial\kappa}\frac{\partial^2}{\partial t^2}h(t,\kappa)\sim|t|^{-\lambda},\label{dgamma}
\ee
with the same exponent $\lambda = \phi - f$ as in Eq.\ \eqref{susceptibility}.

\textit{Computational model} \textemdash\,In order to test the scaling relations derived above, we study two complementary models of fiber networks: triangular lattice-based networks and jammed packing-derived networks. Our triangular networks consist of fibers of length $W$ arranged on a periodic triangular lattice with lattice spacing $l_0=1$, with freely-hinging crosslinks attaching overlapping fibers. To avoid singular mechanical behavior deriving from infinitely long fibers, we initially cut a single randomly chosen bond on each fiber, yielding an initial network coordination number $z$ approaching 6 from below with increasing system size \cite{Das2007, Broedersz2011NatPhys}. 
We prepare packing-derived networks by populating a periodic square unit cell of side length $W$ with $N=W^2$ randomly placed, bidisperse disks with soft repulsive interactions and with a ratio of radii of $1\mathpunct{:}1.4$.  The disks are uniformly expanded until precisely the point at which the system exhibits a finite bulk modulus, after which a contact network excluding rattlers is generated \cite{VanHecke2010, Ohern2003, Dagois-Bohy2012}. Sufficiently large networks prepared using this protocol have an initial connectivity $z\approx z_c$ \cite{Goodrich2012}.

For both network structures, we reduce $z$ to a value below the isostatic threshold by bond dilution and removal of dangling ends and clusters that do not contribute to the network's bulk mechanical response \cite{SupplementaryMaterials}. 
We use a purely random dilution process, in contrast with special cutting protocols that have been used previously to suppress variation in local connectivity and promote mean-field behavior \cite{Wyart2008, Tighe2012}. We perform simulations on ensembles of at least $50$ network realizations each for triangular networks and $30$ networks each for packing-derived networks, both with $z = 3.3$. Unless otherwise stated, we use triangular networks of size $W = 140$ and packing-derived networks of size $W = 120$. Examples of the final disordered network structures are shown in Fig.\ \ref{fig1}b-c. 

We treat each bond as a Hookean spring with 1D modulus $\mu$, such that the total contribution of stretching to the network energy is
\be
\mathcal{H}_S = \frac{\mu}{2}\sum_{\langle ij \rangle}\frac{\left(l_{ij}-l_{ij,0}\right)^2}{l_{ij,0}}
\ee
in which $l_{ij}$ and $l_{ij,0}$ are the length and rest length, respectively, of the bond connecting nodes $i$ and $j$. Bending interactions are included between pairs of nearest-neighbor bonds, which are treated as angular springs with bending modulus $\kappa$. For triangular networks, bending interactions are only considered between pairs of bonds along each fiber, which are initially collinear, whereas for packing-derived networks we account for bending interactions between all nearest-neighbor bonds, as in typical bond-bending networks. Thus, the total contribution of bending to the network energy is
\be
\mathcal{H}_B = \frac{\kappa}{2}\sum_{\langle ijk \rangle}\frac{\left(\theta_{ijk}-\theta_{ijk,0}\right)^2}{l_{ijk,0}}
\ee
in which $\theta_{ijk}$ and $\theta_{ijk,0}$ are the angle and rest angle, respectively, between bonds $ij$ and $jk$, and $l_{ijk,0}=(l_{ij,0}+l_{jk,0})/2$. As we are interested only in the relative contributions of bending and stretching, we set $\mu=1$ and vary the dimensionless bending stiffness  $\kappa$ \cite{Note_on_kappa}.

\begin{figure}[ht]
\includegraphics[width=1.0\columnwidth]{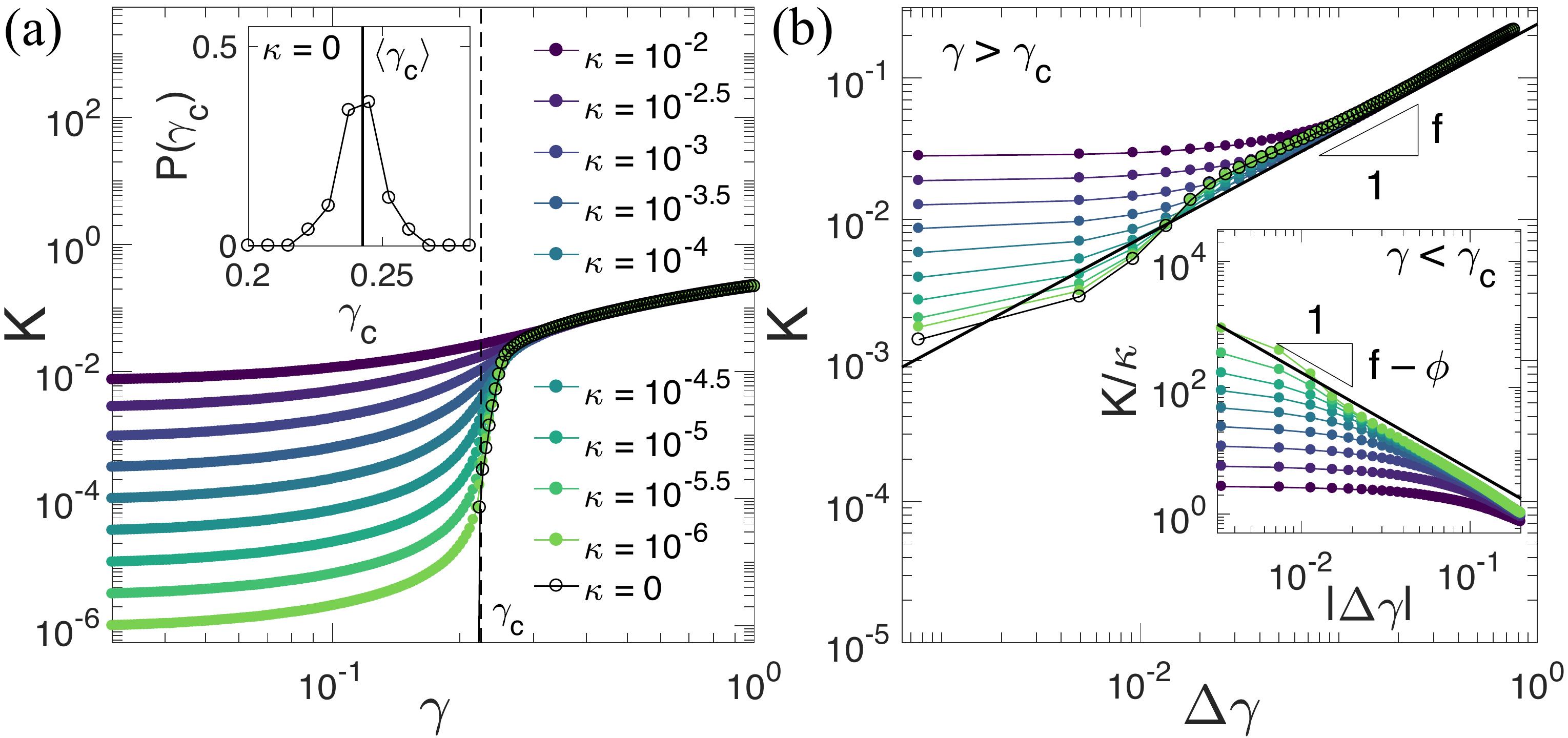}
\caption{\label{fig2} (a) Differential shear modulus $K$ vs. shear strain for triangular networks of size $W = 140$ and connectivity $z = 3.3$, with varying reduced bending stiffness $\kappa$. The dashed line indicates the observed critical strain $\gamma_c$ for the ensemble. The inset shows the probability distribution for the measured $\gamma_c$ values for 50 individual network samples with $\kappa = 0$. (b) For $\gamma>\gamma_c$ and with decreasing $\kappa$, $K$ converges to the form $K\sim\left|\gamma-\gamma_c\right|^f$, with $f = 0.73 \pm 0.04$. These data are for the same networks as in (a). Inset: In the low-$\kappa$ limit and below $\gamma_c$, $K/\kappa$ converges to a power law in $\left|\Delta\gamma\right|$ with exponent $f-\phi \approx -1.5$.  
}
\end{figure}

We apply incremental simple shear strain steps using Lees-Edwards periodic boundary conditions \cite{Lees1972}, minimizing the total network energy $\mathcal{H}=\mathcal{H}_S + \mathcal{H}_B$ at each step using the FIRE algorithm \cite{Bitzek}. We compute the stress tensor as a function of strain as
\be
\sigma_{\alpha \beta} = \frac{1}{2A}\sum_{\langle ij \rangle} f_{ij,\alpha} r_{ij,\beta}
\ee
in which $\mathbf{r}_{ij}$ is the vector connecting nodes $i$ and $j$, $\mathbf{f}_{ij}$ is the force exerted by node $j$ on node $i$ \cite{DoiEdwards}, and $A$ is the area of the periodic box containing the network. For the triangular lattice, $A=(\sqrt{3}/2)W^2$, and for packing-derived networks, $A = W^2$. 
The differential shear modulus $K$ is computed as $K = \partial\sigma_{xy}/\partial\gamma$. To symmetrize $K$, we shear each network in both the $\gamma>0$ and $\gamma<0$ directions. Figure \ref{fig2}a shows $K(\gamma)$ for triangular networks with varying bending rigidity. 

\textit{Results} \textemdash\ First, we consider the scaling of $K$ as a function of strain near $\gamma_c$.  We determine $\gamma_c$ for individual samples as the strain corresponding to the onset of finite $K$ in the CF ($\kappa = 0$) limit, and utilize the mean of the resulting distribution, $\langle \gamma_c\rangle$, for our scaling analysis. The $\gamma_c$ distribution for triangular networks of size $W=140$ is shown in Figure \ref{fig2}a. We observe that with increasing system size, the width of the $\gamma_c$ distribution decreases \cite{SupplementaryMaterials}. 

\begin{figure}[ht]
\includegraphics[width=1.0\columnwidth]{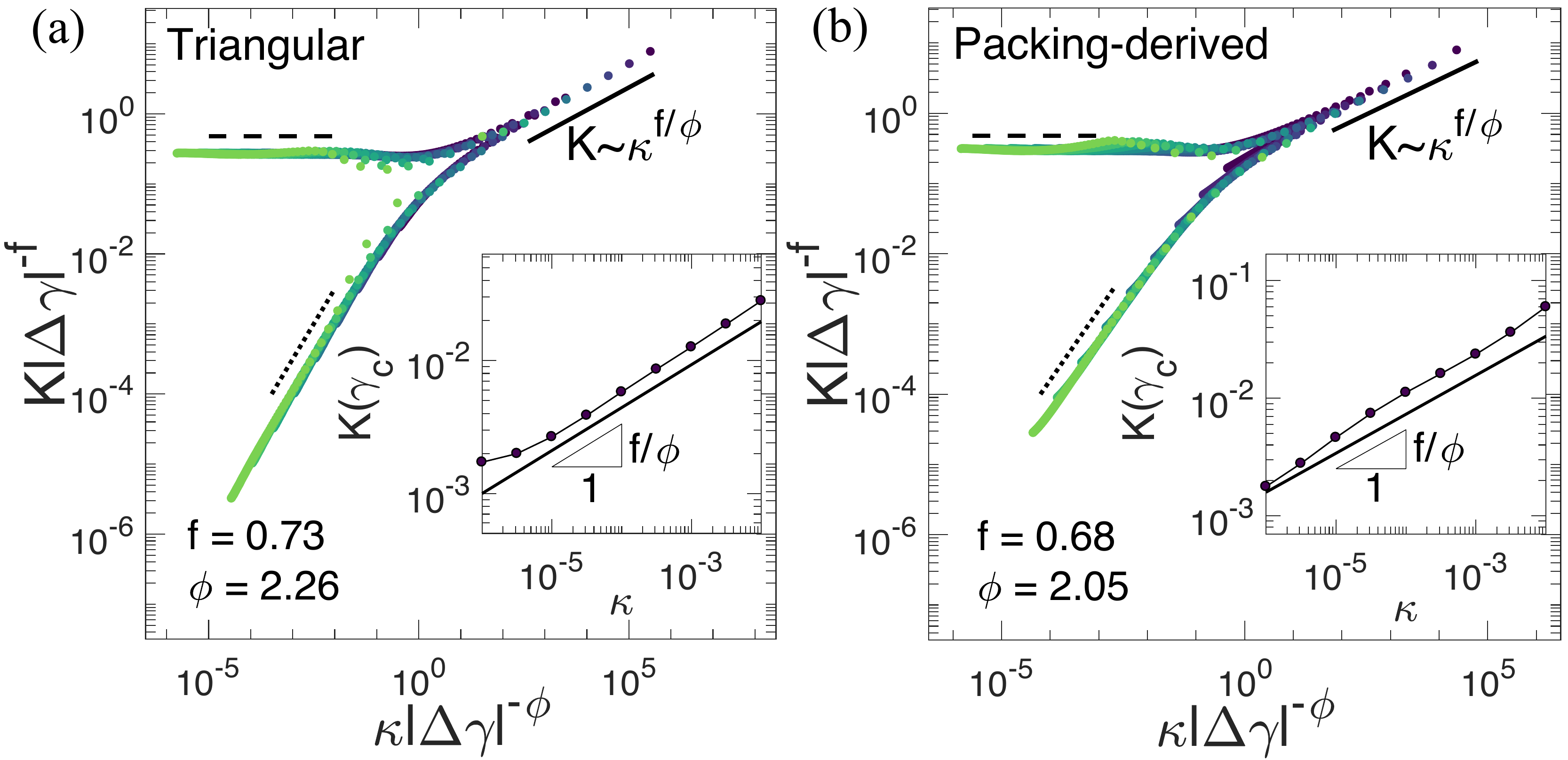}
\caption{\label{fig3} Plotting the $K$ vs. $|\Delta\gamma|$ data for both (a) triangular networks and (b) packing-derived networks according to the Widom-like scaling predicted by Eq.\ \ref{Widom}, and using the values of $f$ and $\phi$ determined previously, yields a successful collapse for both systems. The dashed lines in each panel have slope $0$ and the dotted lines have slope $1$. Insets: At the critical strain, $K\sim\kappa^{f/\phi}$.}
\end{figure}

We then determine $f$ from $K\sim|\gamma-\gamma_c|^f$ in the low-$\kappa$ limit. We obtain a distribution of estimated $f$ values using sample-specific $K$ curves and $\gamma_c$ values for networks with $\kappa = 0$, yielding an estimate of $f = 0.73 \pm 0.04$ for triangular networks, as shown in Fig.\ \ref{fig2}b with decreasing $\kappa$.
Similarly, for packing-derived networks we find $f = 0.68 \pm 0.04$ \cite{SupplementaryMaterials}.  We then estimate $\phi$ by averaging values computed from two separate scaling predictions, as follows. For $\gamma<\gamma_c$, we show the results for Eq.\ \eqref{susceptibility} in the inset to Fig.\ \ref{fig2}b. We also note that continuity of $K$ as a function of strain near $\gamma_c$ requires that $\mathcal{G}_\pm(s)\sim h_{2,0}(\pm1,s)\sim s^{f/\phi}$ for large $s$. Thus, $K(\gamma_c)\sim\kappa^{f/\phi}$, as shown in the insets of Fig.\ 3a-b. Averaging the $\phi$ values computed from these corresponding fits, together with our previously determined value for $f$, we estimate $\phi = 2.26 \pm 0.09$ for triangular networks and $\phi = 2.05 \pm 0.08$ for packing-derived networks. These values of $f$ and $\phi$ are used in Figs.\ \ref{fig3}a-b, which demonstrate the collapse according to Eq.\ \eqref{Widom}.

We compute the nonaffine fluctuations $\delta\Gamma$ as 
\be
\delta\Gamma = \frac{1}{Nl_c^2\delta\gamma^2}\sum_i{\lVert \delta \mathbf{u}_i^\mathrm{NA}\rVert^2}
\ee
in which $N$ is the number of nodes, $l_c$ is the average bond length, and $\delta\mathbf{u}_i^\mathrm{NA} = \delta\mathbf{u}_i -\delta \mathbf{u}_i^\mathrm{A}$ is the nonaffine component of the displacement of node $i$ due to the imposed incremental strain $\delta\gamma$.
Plotting $\delta\Gamma$ vs.\ $\gamma - \gamma_c$ in Fig.\ 4a, we observe agreement with the scaling predicted from Eq.\ \eqref{dgamma} using the $f$ and $\phi$ values determined above. Importantly, as predicted, the corresponding critical exponent $\lambda=\phi-f$ is the same as for Eq.\ \eqref{susceptibility}, with $\lambda\simeq1.5$ \cite{RensPrivateCommunications}. Further, we observe that near $\gamma_c$, the expected scaling $\delta\Gamma(\gamma_c)\sim\kappa^{f/\phi - 1}$ appears to be satisfied \cite{SupplementaryMaterials}.

\begin{figure}[ht]
\includegraphics[width=1.0\columnwidth]{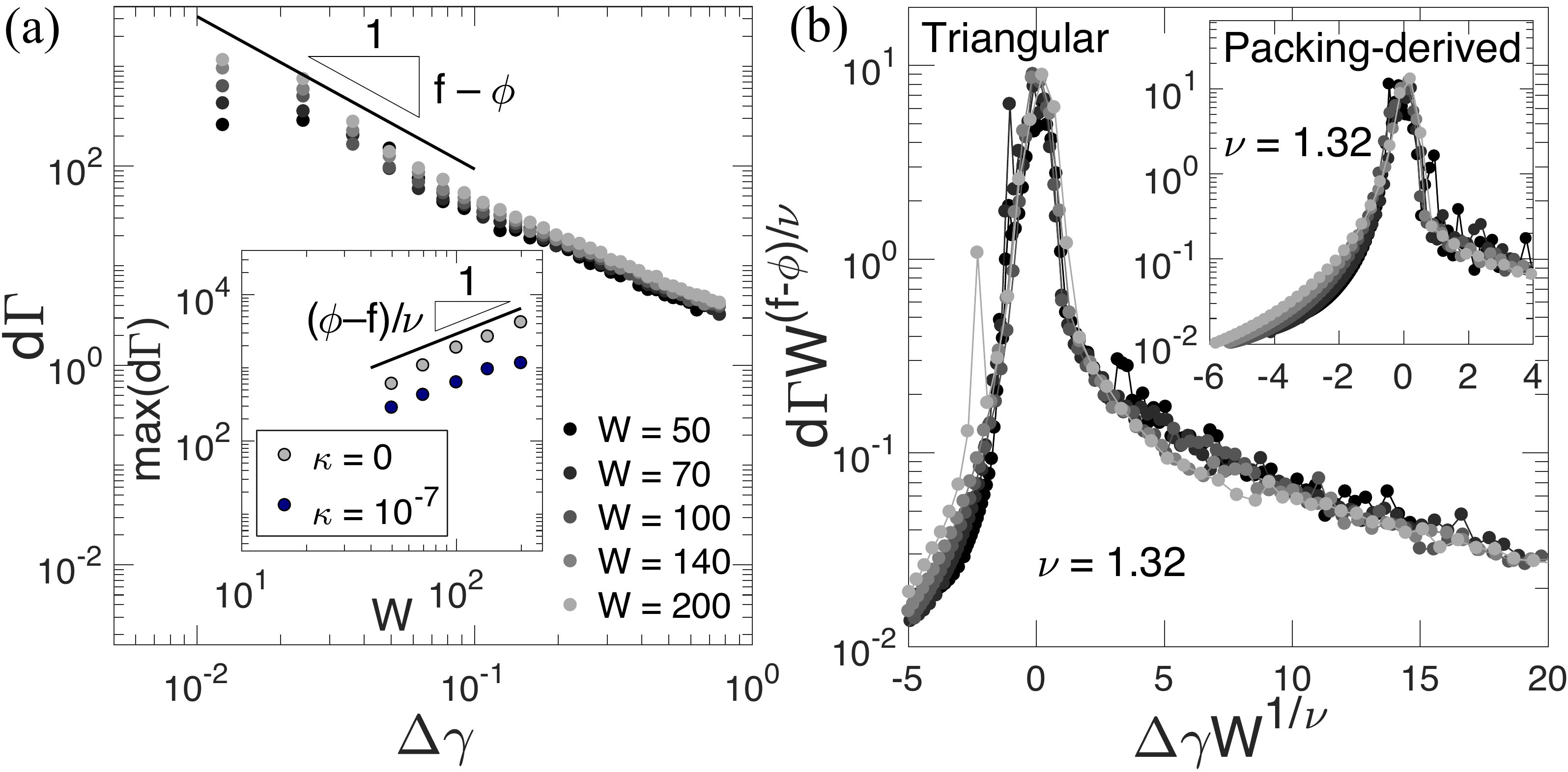}
\caption{\label{fig4} (a) Near the critical strain, the nonaffinity scales as $\delta\Gamma\sim |\Delta\gamma|^{f-\phi}$. These data correspond to triangular networks with $\kappa = 10^{-7}$ and $z = 3.3$, with varying system size. Inset: Nonaffine fluctuations are limited by the system size. For small or zero $\kappa$, the maximum of $\delta\Gamma$ scales as $ \mathrm{max}(\delta\Gamma)\sim W^{(\phi-f)/\nu}$, with $\nu = 1.3 \pm 0.2$. (b) Plots of $\delta\Gamma/W^{(\phi-f)/\nu}$ vs. $(\gamma-\gamma_c)W^{1/\nu}$ for triangular networks and (inset) packing-derived networks for with $\kappa = 0$ demonstrate successful scaling collapse using the $f$ and $\phi$ values determined earlier, with $\nu$ values determined from the scaling relation.}
\end{figure}

It is apparent from Fig.\ 4a that the divergence of the fluctuations is suppressed by finite-size effects. This is consistent with a diverging correlation length $\xi\sim|t|^{-\nu}$. 
Critical effects such as the divergence of $\delta\Gamma$ will be limited as the correlation length becomes comparable to the system size $W$, corresponding to a value of $t\sim t_W=W^{-1/\nu}$. Thus, the maximum value of $\delta\Gamma$ increases as $\delta\Gamma\sim W^{(\phi-f)/\nu}$ (Fig.\ \ref{fig4}a inset). From least-squares fits to this scaling for both triangular and packing-derived networks with $\kappa = 0$ and $\kappa = 10^{-7}$, combined with our estimates for $\phi$ and $f$, we determine that $\nu = 1.3 \pm 0.2$ for both systems.  
We then verify that this leads to a scaling collapse in a plot of $\delta\Gamma W^{(f-\phi)/\nu}$ vs $tW^{1/\nu}$ for both systems with $\kappa = 0$, as shown in Fig.\ 4b, and with finite $\kappa$ \cite{SupplementaryMaterials}. This finite-size scaling is consistent with the (hyperscaling-like) relation $f=2\nu-2$ in 2D from Eq.\ \eqref{strain_exponents}. 

\textit{Near the isostatic point} \textemdash\,  For networks near the isostatic transition at $z=z_c$, we can define a dimensionless distance $\Delta=z-z_c$ from the isostatic point and 
let $h(\gamma,\kappa,\Delta)$ be the Hamiltonian or elastic energy per unit cell. 
At the isostatic point, since $\gamma_c=0$, $t$ above reduces to the strain $\gamma$. 
Assuming the system becomes critical as $\gamma, \kappa, \Delta\rightarrow0$, we can follow a similar real-space renormalization procedure as above, resulting in
\be
h(\gamma,\kappa,\Delta)=L^{-d}h(\gamma L^x,\kappa L^y,\Delta L^w).\label{ScalingIso}
\ee 
Although the exponents $x$, $y$, and $w$ at the isostatic point can be assumed to be positive, we do not necessarily assume the same values of the exponents $x$ and $y$ as determined for the strain-controlled transition. 
We can again determine the stress $\sigma$ and stiffness $K$ as in Eq.\ \eqref{Kdef}.
By letting $L=|\Delta|^{-1/w}$, we again identify the correlation length exponent $\nu'=1/w$ and find 
\be
K\sim |\Delta|^{f'}h_{2,0,0}(0,\kappa/|t|^{\phi'},\pm1),\label{WidomIso}
\ee
where
\be
f'=(d-2x)\nu',\quad\phi'=y\nu'.\label{ScalingRelIso}
\ee
Moreover, following similar arguments as above, it can be shown that $\delta\Gamma\sim|\Delta|^{-\lambda'}$, where $\lambda'=\phi'-f'$ \cite{SupplementaryMaterials,ChaseThesis}, consistent with the values $f'\simeq1.4\pm0.1$, $\phi'\simeq3.0\pm0.2$, $\nu'\simeq1.4\pm0.2$, and $\lambda'\simeq2.2\pm0.4$ reported in Ref.\ \cite{Broedersz2011NatPhys}. 
While our approach uses the elastic energy, it is interesting to note that prior work on rigidity percolation has suggested the use of the number of floppy modes as a free energy \cite{jacobs1995generic}.

\textit{Conclusion} \textemdash\, The scaling theory and relations derived here for the strain- and connectivity-controlled rigidity transitions in athermal fiber networks are consistent with our numerical results, as well as those reported previously observations near the isostatic point \cite{Broedersz2011NatPhys,Das2012,Feng2016}. 
Interestingly, for the subisostatic, strain-controlled transition, we observe that simulations of both triangular networks and packing-derived networks exhibit consistent non-mean-field exponents. 
This, together with agreement with the hyperscaling relation in Eq.\ \eqref{strain_exponents} suggest that the upper critical dimension for fiber networks is $d_u>2$, in contrast with jammed networks at the isostatic point \cite{Goodrich30082016}. 
Our observations, combined with prior scaling with similar exponents for alternate subisostatic network structures, including 2D and 3D phantom networks, branched (honeycomb) networks, and Mikado networks \cite{sharma2016strain, Rens}, suggest that non-mean-field behavior might be ubiquitous in subisostatic networks, irrespective of the local network structure.
Interestingly, the hyperscaling relation in Eq.\ \eqref{strain_exponents}, together with the observation that $f>0$, suggests that fiber networks satisfy the Harris criterion \cite{Harris1974}, which would imply that such networks should be insensitive to disorder.
Further work will be needed to test this hypothesis, as well as the scaling relations derived here in 3D. 
This will likely require finite-size scaling to identify the correlation length exponent, which has not been possible to date in 3D. 

\vspace{3mm}

This work was supported in part by the National Science Foundation Center for Theoretical Biological Physics (Grant PHY-1427654).
The authors acknowledge helpful discussions with Andrea Liu, Tom Lubensky and  Michael Rubinstein, as well as discussions with Edan Lerner and Robbie Rens on central force packing-derived networks.

\bibliography{Refs}{}

\begin{thebibliography}{34}
\bibitem [{Not()}]{Note_on_kappa}%
  \BibitemOpen
  \href@noop {} {}\bibinfo {note} {Note that $\kappa$ refers to the
  dimensionless bending stiffness. $\kappa\equiv\kappa_b/(\mu l_c^2)$, in which
  $\kappa_b$ is the bending rigidity used in the simulations and $l_c$ is the
  average bond length in the unstrained network, $l_c = \langle l_{ij,0}
  \rangle$. This constitutes only a minor adjustment since $l_c=l_0=1$ for the
  triangular networks and $l_c\approx1.03$ for the packing-derived
  networks.}\BibitemShut {Stop}%
\end{thebibliography}

\end{document}